\documentclass{article}
\usepackage[utf8]{inputenc}
\usepackage{authblk}
\usepackage{hyperref}
\usepackage{bm}

\title {Einstein, Free Creations, and His Worldly Cloister}
\author{Galina Weinstein}
\affil{\normalsize The Department of Philosophy, University of Haifa, Israel.} 

\begin{document}

\maketitle

\begin{abstract}
This paper examines \emph{Free Creations of the Human Mind: The Worlds of Albert Einstein} by Diana Kormos Buchwald and Michael D. Gordin. The authors seek to dispel the long-standing myths of Einstein as the "lone genius" of Bern and the "stubborn sage" of Princeton, drawing on newly uncovered archival materials to illuminate his intellectual networks and collaborative engagements. By exploring the authors' reasoning, this paper engages with their interpretations, highlighting the strengths of their archival revelations and areas where alternative perspectives may enrich the understanding of Einstein's intellectual development.
\end{abstract}

\section{Introduction}

This paper examines \emph{Free Creations of the Human Mind: The Worlds of Albert Einstein}, authored by Diana Kormos Buchwald and Michael D. Gordin. Drawing upon newly uncovered archival material, Buchwald and Gordin seek to dispel long-standing myths surrounding Einstein's intellectual legacy, offering fresh insights into his scientific contributions and philosophical reflections. Through carefully analyzing their interpretive framework, this paper highlights the strengths of their archival discoveries but identifies alternative historiographical perspectives. In doing so, it aims to contribute to ongoing scholarly debates about the complexities and nuances of Einstein's thought, while foregrounding the broader implications of his work for the history and philosophy of science.

\section{The Political and Social Resonance of General Relativity}

The book's prologue presents Albert Einstein's first visit to the United States in 1921 as a profound intersection between scientific achievement and political activism. Against the backdrop of international acclaim for his general theory of relativity, confirmed by Arthur Eddington's 1919 eclipse expedition, Einstein emerges not merely as a physicist but as a global figure whose intellectual contributions transcend the boundaries of science. The narrative intricately weaves Einstein's role as a scientific revolutionary with his political engagements, particularly his support for Zionism and Jewish advocacy. His arrival in New York, framed by the \emph{New York Times} as a spectacle of intellectual brilliance, is portrayed as a confluence of scientific triumph and cultural significance \cite{BG}.

Einstein's visit is contextualized within the broader political landscape of post-World War I Europe, marked by rising anti-Semitism and political hostility. The prologue underscores how Einstein's outspoken criticism of anti-Semitic attacks in Berlin, coupled with his support for Eastern European Jewish immigrants, positioned him as both a scientific icon and a political advocate. This duality is further exemplified by his involvement with the Zionist movement and his support for the establishment of a Hebrew University, an initiative reflecting his commitment to education and Jewish autonomy in response to European exclusion.

Buchwald and Gordin also emphasize the symbolic weight of Eddington's confirmation of relativity, not only as a validation of Einstein's scientific work but as a beacon of internationalism amid political division. In a period when German scientists were ostracized from global scientific discourse, Einstein's theory served as a bridge, reconnecting German intellectualism with the Anglo-Saxon scientific community. His Princeton lectures, later published as \emph{The Meaning of Relativity}, are depicted as a crystallization of his theoretical work, marking his transition from European isolation to American scientific prominence \cite{BG}.

Notably, the narrative adopts a socio-political lens rather than a purely scientific one. It situates general relativity within Einstein's broader struggles against political hostility and cultural exclusion, framing his scientific success as inseparable from his political advocacy. Absent from this account, however, is an exploration of the intellectual journey that led Einstein to general relativity. Instead, the focus remains on the societal ramifications of his work, particularly its reception in politically charged contexts.
Thus, the narrative serves as a prelude to understanding Einstein not just as the solitary genius of popular mythology, but as an emblem of scientific internationalism, political advocacy, and cultural resistance. His contributions to general relativity are presented less as isolated intellectual triumphs and more as intertwined with the political and social currents of his time.

\section{The Myth of Two Einsteins and the Reality of Continuity}

Albert Einstein's life was marked by significant geographical and intellectual transitions that defined his scientific achievements and public image. His early years were notably peripatetic: born in Ulm, Germany, he moved to Munich, then to Italy (Pavia and Milan), and several cities in Switzerland. However, after a period of mobility, his life stabilized in two major intellectual centers: Berlin (1914–1933) and Princeton (1933–1955), which framed the most transformative periods of his career \cite{BG}. 

In the book's first chapter, Buchwald and Gordin undertake a demythologizing project by strategically concentrating on two pivotal periods in Einstein's life: his years in Bern (1902–1909) and Princeton. This selective framing, however, deliberately omits the Berlin period (1914–1933), despite it being the era of Einstein's most profound scientific achievements, including the formulation of general relativity. 
The authors justify this choice by arguing that these locales—one (Bern) marking the beginning of his career and the other (Princeton) its end—had an outsized impact on Einstein's life and public image. By isolating these two periods, the book attempts to demythologize the familiar "Young Genius" and "Old Sage" binaries by emphasizing intellectual and collaborative continuities \cite{BG}.

According to Buchwald and Gordin, while Berlin represented the pinnacle of his scientific contributions, the earlier years in Bern and the later years in Princeton symbolized two distinct public images of Einstein \cite{BG}. 

\subsection{Bern: The Networked Genius} 

This phase established the myth of the lone genius who reshaped scientific thought from the confines of a patent office. 
The romanticized image of Einstein in Bern portrays him as the quintessential lone genius—a patent clerk detached from academia who single-handedly revolutionized physics in 1905 with groundbreaking papers on the photoelectric effect, Brownian motion, and special relativity. This idealization perpetuates the notion of solitary brilliance—a maverick outsider transcending institutional constraints to achieve scientific immortality.

However, this narrative, while compelling, is reductive. It neglects the intellectual networks that shaped Einstein’s thinking. Although the annus mirabilis papers of 1905 were authored solely by Einstein, his intellectual environment was anything but isolated. 

In Bern, Einstein was deeply engaged in collaborative discourse with close friends and colleagues such as Conrad Habicht, Michele Besso, Joseph Sauter, and Lucian Chavan. The Olympia Academy, a reading group he formed with Habicht and Maurice Solovine, was an intellectual incubator with rigorous discussion of scientific and philosophical ideas. Texts by Ernst Mach, David Hume, and Henri Poincaré were foundational to Einstein’s evolving theoretical perspectives \cite{BG}.

Einstein’s role at the Swiss Patent Office is also misunderstood. Far from being a mere clerk, his work involved evaluating technical innovations, familiarizing him with cutting-edge electrical technologies, and patent law. This practical engagement with technology influenced his thinking about time, synchronization, and relativity. Thus, the Bern period was not one of isolation but of intellectual community, philosophical reflection, and practical engagement with the mechanics of innovation.

This understanding challenges the myth of isolated brilliance—epitomized by the caricature of Einstein supposedly declaring, "I will little tink," (as he would later say in his imperfect English at Princeton) before pulling out his pen to scribble his papers \cite{Hoff}. Instead, it replaces this myth with the concept of a networked genius whose ideas were nurtured through dialogue and collective exploration, à la Mara Beller \cite{Bel}. During this time, Einstein’s foundational ideas on space, time, and light coalesced into the 1905 revolutionary papers. These developments did not emerge in a vacuum but were the products of intense intellectual collaboration and engagement with contemporary scientific debates.

\subsection{Princeton: The Collaborative Sage}

By contrast, Einstein's time in Princeton marked his transformation into a global intellectual and humanitarian icon. His appearance evolved into the archetype of the white-haired sage, dressed in idiosyncratic fashion, symbolizing wisdom and moral conviction. During these years, he became a vocal advocate for pacifism, Zionism, and opposition to Nazism. However, the physics community largely dismissed his scientific work on unified field theory and his challenges to quantum mechanics, contributing to the myth of the aging sage disconnected from contemporary science \cite{BG}.

Buchwald and Gordin argue that popular narratives often reduce Einstein to two distinct figures: the young Einstein of Bern, the revolutionary scientist who transformed physics through sheer intellectual brilliance.
The old Einstein of Princeton—the humanitarian and public intellectual, whose scientific ideas were perceived as relics of the past.
While compelling, this binary division overlooks the intellectual continuity that spanned his entire life. Despite geographical and political changes, Einstein's intellectual mission remained remarkably consistent. His scientific exploration, philosophical engagement, and struggles with institutional barriers persisted from his youthful days in Bern through his celebrated years in Princeton. Although differently perceived, his creativity and commitment to fundamental questions in physics linked these two phases of his life in profound ways \cite{BG}.

\subsection{Einstein's Marginalized Intellectual Life}

Buchwald and Gordin strive to demythologize the image of Einstein as a "lone genius" by highlighting his interactions with friends, colleagues, and assistants during his formative years in Bern and later at Princeton. That is a fair point. However, my concern is that this attempt at demythologization may inadvertently reinforce the myth it aims to dismantle. Through a nuanced examination of Einstein's intellectual environments—from the "worldly cloister" of the Patent Office in Bern to his isolated office at the Institute for Advanced Study in Princeton—a pattern of marginalization emerges, one that mirrors his solitary brilliance across different phases of his career.

Einstein's years in Bern are often romanticized as the time of the "young genius" working in splendid isolation, conjuring revolutionary ideas from the quiet solitude of the Patent Office. The authors, however, seek to challenge this portrayal by introducing the Olympia Academy—Einstein's self-made intellectual circle, where he and his friends, Habicht and Solovine, debated philosophy and science in the evenings. 
Additionally, Einstein found companionship and intellectual engagement with his colleagues, Besso, Chavan, and Sauter, at the Patent Office. This was not solitude; it was a carefully constructed microcosm—a "department of physics" of his own making, far from the bustling corridors of academia.

Yet, while this reframing strips away the myth of complete isolation, it subtly reinforces another: the myth of intellectual self-sufficiency within an institutional vacuum. The Olympia Academy and the small circle at the Patent Office were not replacements for academic networks; they were the makeshift scaffolding of a brilliant mind shut out from the structures of formal academic discourse. Einstein's intellectual sanctuary was, in fact, a response to his exclusion from the institutionalized academic world. His isolation was not entirely self-imposed; it also reflected the barriers to academic acceptance.

When Einstein transitioned to professorships in Zurich and later Berlin, the structure of his intellectual life did not fundamentally change. Although he gained institutional recognition, his mode of work remained insular. Einstein collaborated primarily with a handful of assistants, not large research groups typical of academic settings. His intellectual collaborations remained limited to those directly under his tutelage or assisting with his calculations. His partnership with Marcel Grossmann in Zurich and later with assistants in Berlin mirrored the dynamics of the Patent Office: small, tightly controlled intellectual environments where Einstein's ideas could flourish largely undisturbed by external critique.

In Berlin, despite being surrounded by luminaries such as Max Planck and Walther Nernst, Einstein's work on general relativity remained an isolated pursuit. The field of general relativity itself was marginalized during Einstein's lifetime—an intellectual island scarcely populated by more than a handful of active researchers. His office in Berlin became, much like the Patent Office, a worldly cloister where ideas could germinate in relative isolation, removed from the broader academic community.

Buchwald and Gordin emphasize Einstein's relocation to Princeton as the culmination of his acceptance into the highest echelons of academia. Yet, a closer examination reveals a replication of his previous intellectual solitude. At the Institute for Advanced Study, Einstein was surrounded by brilliant minds—John von Neumann, Kurt Gödel, and others—yet his work remained largely disconnected from the mainstream developments in physics. He was institutionally housed but intellectually isolated. His focus on unified field theory, a pursuit dismissed by many of his contemporaries, was mainly conducted within the confines of his office, supported by a few assistants who mirrored the roles once filled by Besso and Solovine.

Einstein's office at Princeton became a new kind of worldly cloister, an echo of the Patent Office but within the polished walls of academia. Though now institutionally legitimized, his intellectual life in Princeton was as insular as ever. He did not direct large research groups; his assistants worked with him like his colleagues in Bern—an enclave of solitary brilliance masquerading as academic integration.

The authors’ attempt to demythologize Einstein's image as a lone genius unveils a continuity of intellectual marginalization masked by institutional acceptance. From Bern to Princeton, Einstein operated within self-created cloisters that mimicked academic structures without truly integrating into them. This peculiar isolation—self-sustained yet institutionally tolerated—suggests that Einstein's intellectual life was not a story of integration into mainstream academic physics but a perpetuation of his earlier marginalization, albeit within more prestigious walls.

The mirage of inclusion created the illusion of acceptance while Einstein's most groundbreaking ideas, particularly in general relativity, remained at the periphery of mainstream physics until long after his death. In this sense, Buchwald and Gordin’s demythologization inadvertently reinforces the narrative of Einstein as a figure whose intellectual solitude was not entirely overcome by academic recognition but merely transplanted into more gilded surroundings.

\section{Relativity and Intellectual Scaffolding}

Buchwald and Gordin's narrative articulates a historiographical perspective that challenges the mythos of Einstein as a solitary genius who revolutionized physics ex nihilo. Instead, it situates his 1905 formulation of special relativity within a well-established intellectual framework, drawing from the conceptual contributions of Galileo, Lorentz, and Poincaré \cite{BG}. This approach is conceptually aligned with Michel Janssen's metaphor of "scaffolding," which posits that Einstein's theories were constructed upon existing scientific structures, reconfigured through critical engagement rather than radical disjunction \cite{Jan}.

The author's narrative adopts the metaphor of scaffolding to illustrate the intellectual infrastructure underpinning Einstein's breakthrough. For instance, Galileo's principle of relativity is presented as a foundational scaffold dating back to \emph{Dialogue Concerning the Two Chief World Systems} (1632) \cite{Gal}, where inertial motion is shown to be indistinguishable from rest through thought experiments aboard a ship. This conceptual basis remained structurally sound through the developments of classical mechanics and served as Einstein's point of departure when extending the principle to electromagnetism \cite{BG}.

Here, scaffolding is metaphorical, signifying a dynamic conceptual ecosystem rather than mere historical influence. Einstein's role was not that of a solitary innovator erecting theory in a vacuum but rather that of a critical architect who modified, extended, and solidified existing structural elements into a coherent theoretical framework. This intellectual reconfiguration demystifies the notion of the lone genius, portraying Einstein instead as a synthesizer who adeptly reworked established ideas.

\subsection{Lorentz and Poincaré: Building Blocks in Einstein's Conceptual Architecture}

The book's narrative emphasizes that "eminent theorists of his day, including Hendrik Lorentz and Henri Poincaré, failed to push as far or as rigorously as the Swiss patent clerk," signaling that Einstein's work was deeply scaffolded by their contributions \cite{BG}. This perspective aligns with a historiographical interpretation that positions Einstein's work on special relativity as a synthesis and extension of foundational contributions by Lorentz and Poincaré, rather than as an isolated intellectual leap. His genius resided in reassembling these established principles into a coherent and transformative theoretical architecture, rather than generating entirely novel concepts ex nihilo.

Yet, the role of Poincaré in this scaffolding remains historiographically ambiguous. While Poincaré's work on clock synchronization and the principle of relativity is often cited as an intellectual precursor to Einstein's special relativity, there is scant primary evidence to confirm that Einstein engaged directly with Poincaré's writings before 1905. This ambiguity introduces a historiographical gray zone that complicates claims of direct intellectual inheritance. 
Poincaré's discussions of simultaneity and clock synchronization prefigure key aspects of Einstein's 1905 paper. However, primary historical evidence that Einstein engaged with it before 1905 is notably absent. The connection is often inferred circumstantially.
This historiographical ambiguity introduces what might be termed an "argument from silence." 

\subsection{The Ether Hypothesis: Dismantling Outdated Scaffolding}

Buchwald and Gordin suggest that Einstein's 1905 relativity paper appeared anomalous to his contemporaries due to its lack of citations \cite{BG}. However, there is no historical evidence supporting this claim. Indeed, Einstein's paper did not cite anyone beyond a brief acknowledgment of Besso. The perception of Einstein's work as disconnected from empirical and theoretical foundations is a retrospective interpretation, not a documented reaction from his peers.

The authors further assert that it is unclear whether Einstein knew of the Michelson-Morley experiment when he wrote his 1905 paper, noting that he did not cite it \cite{BG}. Whether Einstein was aware of the Michelson-Morley experiment before 1905 remains contested. Traditional accounts cast the null result as a catalyst for special relativity, yet Einstein's reflections were inconsistent. At times, he acknowledged its influence; at others, he minimized its relevance. Historian Gerald Holton noted that Einstein rarely mentioned Michelson when discussing his theory's origins, instead highlighting Fizeau's experiment and stellar aberration \cite{Hol}. In later correspondence, Einstein expressed uncertainty about when he first learned of Michelson's findings, suggesting they did not consciously shape his thinking \cite{Shan}. He maintained that his path to special relativity was driven by heuristic principles, not ether-drift experiments. Thus, despite its symbolic association with relativity, Einstein's testimony suggests his conceptual leap was rooted in theoretical reasoning rather than direct experimental influence.

One of the narrative's central examples of Einstein's intellectual restructuring is his rejection of the ether hypothesis. For decades, the ether served as the assumed medium for electromagnetic propagation, and its existence was considered essential for transmitting light waves. The Michelson–Morley experiment's null result cast doubt on this assumption, yet the conceptual structure of the ether remained largely intact within scientific discourse.
Einstein's 1905 critical intervention was the decisive dismantling of the ether as a necessary construct in physical theory. While traditionally regarded as a bold act of theoretical destruction, Buchwald and Gordin frame this move as a refinement that echoes Janssen's metaphorical scaffolding \cite{BG}. In this view, Einstein's contribution was not merely the elimination of the ether but a strategic clearing away of superfluous elements to reveal the streamlined architecture of special relativity.

Against this backdrop, the authors cite Einstein's 1920 lecture "Ether and the Theory of Relativity," delivered at the University of Leiden, as evidence of the ether being "not entirely dead, however, not even for Einstein" \cite{BG}. However, the historical context complicates this interpretation. Before the lecture, Einstein wrote to Lorentz, "If you tell me what I should lecture on, I shall adjust myself with pleasure to the need. I shall hold the inaugural lecture you mentioned on the ether" \cite{CPAE9}, Doc. 256. This deference suggests that Einstein's reintroduction of "ether" was partly a gesture of respect for Lorentz, who had never entirely abandoned the concept. Indeed, the "ether" Einstein described in Leiden was purely metaphorical, referring to the geometric properties of spacetime in general relativity, devoid of mechanical or particulate qualities.

Thus, Einstein's reference to "ether" in 1920 did not signify a reconstruction of the discarded scaffolding but rather a rearticulation within the geometric formalism of relativity. His theoretical restructuring from 1905 remained intact, with the 1920 lecture representing a linguistic adaptation rather than a philosophical retreat.

\subsection{General Relativity, gravitational waves, and cosmology}

The book's treatment of general relativity is notably more succinct and analytically focused than its coverage of special relativity. It primarily emphasizes the conceptual development of the theory, tracing Einstein's “happiest thought” of 1907, where he formulated the equivalence principle, to the final presentation of the field equations in 1915 \cite{BG}. While it acknowledges the collaboration with Marcel Grossmann and briefly touches upon the 1913 "Entwurf" theory, the text presents Einstein's path as a linear progression. I recognize the challenge of encapsulating Einstein's multifaceted legacy as a physicist, political activist, and philosopher within 120 pages. However, I find the limited emphasis on the intricate development of general relativity somewhat disconcerting.

The book's narrative contrasts sharply with rich historical accounts that depict Einstein's struggle with tensor calculus, recursive corrections, and intellectual debates with contemporaries such as Max Abraham, Gunnar Nordström, Tullio Levi-Civita, David Hilbert, and Willem de Sitter. The narrative minimizes the intellectual friction, trial-and-error, and contentious exchanges that characterized Einstein's development of general relativity from 1912 to 1916, smoothing over his theoretical achievement's complex, non-linear nature.

Even after finalizing the theory in 1915–1916 and presenting gravitational waves and his static cosmological model, Einstein remained embroiled in intense debates with prominent figures like Levi-Civita, Nordström, de Sitter, and others. These interactions were marked by deep disagreements, conceptual challenges, and contrasting interpretations of his theory. The narrative's omission of this turbulent intellectual period inadvertently perpetuates the notion of a smooth, unchallenged path to general relativity, sidelining the contentious discourse that shaped its maturation.

\section{Quantum Debates and Experimentation}

\subsection{The Compton Effect}

I am not entirely convinced by Buchwald and Gordin's decision to allocate disproportionate emphasis to Einstein's contributions to quantum theory, particularly the Compton effect, at the expense of a more comprehensive exploration of general relativity. 
The book presents general relativity as merely one accomplishment within the broader trajectory of Einstein's scientific contributions, rather than recognizing it as Einstein's monumental achievement. Although it acknowledges Einstein's "happiest thought"—the equivalence principle—and his collaborative efforts in mastering tensor calculus, the text falls short of capturing the profound intellectual and conceptual upheaval that general relativity introduced. This portrayal diverges sharply from Einstein's valuation of his work, as evidenced by the extensive correspondence and numerous papers dedicated to general relativity in the \emph{Collected Papers of Albert Einstein}.
In contrast, the Compton effect—a significant but relatively narrower contribution to quantum mechanics—is discussed in a manner that almost parallels the space given to general relativity, creating an imbalance.

The book emphasizes how Compton's experiment, which demonstrated that X-rays scatter with energy and momentum in a particle-like manner, provided crucial empirical validation for Einstein's light quantum hypothesis. This experimental triumph positioned Einstein as forward-thinking, anticipating pivotal developments in quantum physics even in the face of skepticism. Moreover, the detailed narrative of Compton's experiment challenges the common myth of Einstein as purely theoretical and disconnected from experimental physics. The text recounts Einstein's collaboration with Hans Geiger and Walther Bothe, where they aimed to distinguish between wave and quantum theories of light through direct experimental work. This portrayal is significant because it presents Einstein not merely as a philosophical critic of quantum mechanics but as someone actively engaged in experimental verification—a side of his often underemphasized work \cite{BG}.

In Buchwald and Gordin's narrative, the Compton effect metaphorically bridges two distinct phases in Einstein's engagement with quantum theory. Leading up to this event, the text underscores Einstein's pivotal role in introducing the quantum nature of light, a revolutionary idea that earned him the Nobel Prize for the photoelectric effect. The Compton effect marks a moment of triumph for Einstein's light quantum hypothesis, validating the particle-like behavior of light with undeniable empirical evidence.
However, this moment also symbolizes a turning point. While the Compton effect bridged Einstein's quantum ideas to broader acceptance, it simultaneously marked the boundary after which quantum mechanics evolved in ways that Einstein could not reconcile with his philosophical views. The author's emphasis on the Compton effect highlights its role as a final empirical alignment before Einstein distanced himself from the emerging quantum mechanics frameworks, like Heisenberg's matrix mechanics. Thus, the Compton effect becomes a metaphorical bridge in the narrative, linking Einstein's early quantum triumphs to the new quantum era that would eventually leave him increasingly isolated and critical of its philosophical implications.

Einstein was awarded the Nobel Prize for the photoelectric effect—a recognition that, while significant, fell conspicuously short of honoring his magnum opus, the theory of relativity. Instead, the accolade was granted for what could be considered an opusculum amidst the vast landscape of his scientific achievements. Despite the transformative implications of special and general relativity, the Nobel Committee withheld acknowledgment, citing a lack of conclusive experimental verification and lingering skepticism within the scientific community. From 1905 to 1915, special relativity was met with resistance, as many physicists believed that Lorentz's ether-based theory could equally well explain experiments like Michelson-Morley and Bucherer's mass-velocity measurements. The Committee's 1910 report recommended delaying any award for relativity until further empirical validation could be secured, reflecting the broader scientific conservatism of the time.

By 1915, attention had shifted to general relativity, but Einstein's struggle with generally covariant field equations and the absence of definitive experimental confirmation fueled ongoing skepticism. Attempts to detect gravitational redshift—most notably by Charles Edward St John—yielded inconclusive or negative results, bolstering the Committee's reluctance. Even after Eddington's 1919 solar eclipse expedition famously verified the bending of light, antisemitic critics like Ernst Gehrcke and Philipp Lenard mounted campaigns to discredit Einstein's work, framing relativity as speculative and ideologically suspect. These attacks, coupled with entrenched biases within the Nobel establishment, stymied Einstein's path to recognition for relativity \cite{Pais}.

In a symbolic act of defiance, Einstein chose to deliver his Nobel lecture on relativity, not the photoelectric effect, asserting his conviction that his true legacy lay in general relativity. This gesture underscored the enduring irony of his recognition: the world's most celebrated physicist was honored for what he considered a footnote to his far greater theoretical triumph.

\subsection{Empirical Integrity and the All-or-Nothing Gamble}

According to Buchwald and Gordin, Einstein collaborated with experimentalists like Hans Geiger and Walther Bothe to test the quantum nature of light, even at the risk of disproving his theories. His engagement with the Compton effect is a testament to his readiness to challenge his interpretations if evidence demanded it. He was even prepared to relinquish field theory, a cornerstone of classical physics and his general relativity, if experiments had convincingly refuted its tenets \cite{BG}.

To appreciate Einstein's approach, it is crucial to understand the difference between theories of principle and constructive theories. A theory of principle is built upon broad, foundational axioms that describe the nature of physical reality in a top-down manner. Special relativity, for instance, is based on two unyielding principles: the constancy of the speed of light in all inertial frames of reference and the invariance of physical laws across these frames. General relativity extends this framework through the equivalence principle and general covariance.

Einstein believed one could not simply "tweak" a principle without unraveling the entire logical structure of the theory. So he famously remarked that if even a single experiment contradicted the predictions of relativity, he would abandon the theory entirely.\footnote{Einstein addressed Dayton C. Miller's experiments in "Meine Theorie und Millers Versuche" ("My Theory and Miller's Experiments") \cite{CPAE15}, Doc. 161 (my translation):

\begin{quote}
Suppose you, dear reader, wanted to use this interesting scientific situation to make a bet. In that case, I recommend you bet that Miller's experiments will prove faulty, or that his results have nothing to do with an "ether wind." I would be pretty happy to put my money on that. If Professor Miller's experiments are carried out with the required precision and indeed show a positive effect, I would not hesitate to abandon my theory. For no experiment can ever be replaced by mere experience.     
\end{quote}

This statement reflects Einstein's scientific integrity and confidence in empirical evidence. He was willing to concede his theories if conclusive experimental evidence was presented—a testament to his commitment to scientific principles.
Buchwald and Gordin reference Doc. 161 in their book in the discussion surrounding the ether and Einstein's eventual rejection of it \cite{BG}.} 
In contrast, constructive theories—like the kinetic theory of gases—are assembled piece by piece, modeling interactions step by step. Such theories allow for incremental adjustments.

However, for Einstein, principles not only guided his formulation of theories but also acted as constraints that shaped the boundaries of his theoretical development. In the Zurich Notebook, his work on general relativity was circumscribed by the principles of correspondence (Newtonian limit) and momentum-energy conservation. Furthermore, in his debates on quantum theory, Einstein invented the principle of separability, which restricted his interpretation of quantum phenomena to maintain the independence of spatially separated systems. Einstein saw separability and Heisenberg's uncertainty principle as \emph{guiding constraints} in the same way he saw the principles of conservation and correspondence in his earlier work on general relativity. Thus, Einstein's engagement with these principles reflects not only their heuristic value but also their role as delimiting conditions within his scientific reasoning.

Buchwald and Gordin frame the principles of separability and Heisenberg's uncertainty primarily as philosophical and theoretical contention points for Einstein \cite{BG}. However, they are focused explicitly on Einstein's engagement with quantum mechanics, particularly his objections to the Copenhagen interpretation and his arguments concerning nonseparability and the completeness of quantum mechanics as expressed in the Einstein–Podolsky–Rosen (EPR) paper. They engage in an extensive discussion of the principles of separability and Heisenberg's uncertainty within the confines of quantum theory, exploring their philosophical implications for reality and locality \cite{BG}. But this analysis is not extended to encompass the foundational role of principles in general relativity.

\subsection{Reduction of Complex Dialogues to a Binary Conflict}

The "Einstein–Bohr debate" has long been considered a defining intellectual conflict in the history of quantum mechanics. Popular narratives portray it as a philosophical duel between two titans of physics—Albert Einstein and Niels Bohr—over the nature of quantum reality. However, Buchwald and Gordin suggest that this debate, as commonly understood, is essentially a retrospective construction. This perspective argues that the "debate" was not the sustained, deeply philosophical confrontation it is often depicted to be, but rather an artifact shaped through selective storytelling. 
The authors argue that portraying the Einstein–Bohr debate as a grand philosophical standoff not only oversimplifies the historical reality but also obscures the broader collaborative discourse that characterized the development of quantum mechanics.

They argue that the mythologization of the Einstein–Bohr debate can be traced back to Paul Arthur Schilpp's volume, \emph{Albert Einstein: Philosopher-Scientist} (1949) \cite{Schi}. The book included Einstein's \emph{Autobiographical Notes} \cite{Ein49} and a series of essays by prominent physicists, with Einstein responding to their critiques. Bohr's essay in this volume, "Discussion with Einstein on Epistemological Problems in Atomic Physics" \cite{Bor}, is identified as the origin of the narrative that Einstein and Bohr engaged in an extended, dramatic intellectual clash over the foundations of quantum mechanics. This essay, which took Bohr two years to complete, retroactively framed various scattered discussions between him and Einstein as part of a continuous philosophical duel. Doing so transformed a collective and multi-faceted dialogue among physicists into a mythologized conflict between two individuals. This reframing gave coherence to Bohr's philosophical stance, particularly concerning his ideas of complementarity and indeterminacy, positioning them as methodologically sufficient for describing atomic phenomena \cite{BG}.

The authors argue that one significant consequence of this historical reconstruction is the reduction of the broader, multi-dimensional discourse on quantum mechanics to a binary conflict between Einstein and Bohr. However, in reality, the development of quantum theory involved a wide array of physicists, including Paul Ehrenfest, Wolfgang Pauli, Erwin Schrödinger, and others, who actively contributed to its interpretation and evolution. The mythologized narrative focuses almost exclusively on Einstein and Bohr, erasing these other influential figures' contributions and nuanced positions. This selective focus diminishes the recognition of collaborative dialogues crucial to the theory's advancement.

Bohr's portrayal in Schilpp's volume also artificially solidified his philosophical views as fully formed and dominant during the time of the EPR paper. 
According to Buchwald and Gordin, Bohr's essay reconstituted the scattered and evolving discussions into a structured, coherent narrative that did not entirely reflect the fluidity of quantum debates during that period. This reframing gave the impression that Bohr's interpretation of quantum mechanics was universally accepted and methodologically complete, when there were significant uncertainties and ongoing debates among theorists. As a result, alternative perspectives were marginalized, and the real-time complexities of quantum interpretation were smoothed over to fit a cleaner historical arc \cite{BG}.

\subsection{Einstein and Rupp's Collaboration}

According to Buchwald and Gordin, the mythologized Einstein–Bohr debate overshadows Einstein's practical engagements with quantum mechanics. While conventional accounts often depict Einstein primarily as a theoretical critic of quantum mechanics, the authors suggest his role was more nuanced. 

The authors present Einstein's involvement in quantum mechanics during the late 1920s, focusing on his collaboration with Emil Rupp to investigate the nature of light emission. Contrary to the standard narrative that frames Einstein primarily as a philosophical critic of quantum mechanics, the authors underscore his empirical engagement through experimental work. Einstein's collaboration with Rupp aimed to determine whether excited atoms emit light instantaneously (in quanta) or over a finite duration (in waves), reflecting his continued interest in testing quantum predictions. Despite his skepticism toward the Copenhagen interpretation, Einstein's preference for Erwin Schrödinger's wave mechanics and his participation in experimental investigations suggest a more nuanced role in the quantum debates than is typically acknowledged \cite{BG}.

Buchwald and Gordin illustrate that Einstein remained intellectually active in quantum physics through correspondence with key figures such as Max Born and Schrödinger, even if his critiques did not always surface in published works. His engagement with Rupp is portrayed as part of his broader empirical inquiry, highlighting his willingness to explore the experimental implications of quantum theories, rather than restricting his contributions to theoretical dissent \cite{BG}.

According to the authors, while collaborating with Rupp, Einstein continued to express his skepticism about the completeness of quantum mechanics, most notably in his famous remark to Max Born: “Quantum mechanics is very worthy of respect. But an inner voice tells me this is not the genuine article. The theory delivers much but hardly brings us closer to the Old One’s secret. In any event, I am convinced that He is not playing dice” \cite{CPAE15}, Doc. 426. This remark, which has become emblematic of Einstein’s resistance to probabilistic interpretations, encapsulates his philosophical discomfort with the indeterminacy introduced by quantum mechanics \cite{BG}

In their scientific correspondence, Einstein proposed a series of experimental adjustments to resolve the persistent Doppler broadening issue that interfered with canal ray interference experiments for Rupp. Rupp meticulously followed Einstein's recommendations, including the critical adjustment of rotating the grating $90^0$ and realigning the setup to optimize the visibility of interference fringes. Furthermore, Rupp experimented with different grating distances—specifically $0.1$ mm, $0.2$ mm, and $0.05$ mm—to observe how the interference patterns varied under these conditions. The suggestion by Einstein to remove the lens and align the grating with the canal rays to simplify path differences was successfully implemented, resulting in sharper and more stable fringes.

A primary obstacle in the canal ray experiments was the Doppler broadening caused by the high velocity of the particles in the beam. This effect broadened the spectral lines, disrupting the coherence needed for clear interference. By optimizing the grating distance and refining the beam alignment, Rupp managed to minimize the Doppler-induced spectral shifts. The clear appearance of maxima and minima across significant path differences (up to $45$ cm) strongly indicates that Doppler broadening had been effectively controlled, allowing for stable interference patterns that would otherwise have vanished.

\subsection{Einstein's Enthusiasm}

The success of these interference experiments provided indirect empirical support for Schrödinger's wave mechanics over Heisenberg's matrix mechanics. Rupp's findings demonstrated that the periodic nature of electromagnetic waves must synchronize with atomic periodicity. This observation indirectly aligns with Schrödinger's wave interpretation, where particles are described as wave packets possessing well-defined phase relations. The persistence of interference patterns over long path differences suggested that the particles maintained a continuous wave-like structure, which is a characteristic that contradicts the purely discrete quantum jumps central to Heisenberg's approach. Hence, Rupp's experiments not only addressed the Doppler complications but also provided evidence favoring “Undulationstheorie,” and thus Einstein's and Schrödinger's perspective on quantum mechanics \cite{CPAE15}, Doc. 277.

Einstein's enthusiasm for Rupp's experiments can be directly attributed to their empirically substantiating a wave-based interpretation of quantum phenomena. Einstein had long advocated for realistic and continuous descriptions of particles, placing him in philosophical opposition to Heisenberg and Bohr's probabilistic model. Schrödinger's wave mechanics, with its deterministic and continuous wave functions, was far more compatible with Einstein's vision. Rupp's experimental success, therefore, represented not just a validation of Einstein's theoretical convictions but also a potential challenge to the rapidly growing acceptance of matrix mechanics.

The collaboration was also strategic: Rupp, despite being relatively new to the field, was highly receptive to Einstein's suggestions and capable of conducting precise and well-controlled experiments. Einstein likely saw in Rupp a skilled experimentalist who could bring his theoretical ideas into empirical reality, providing a counter-narrative to the Copenhagen interpretation that was beginning to dominate the field.

Interestingly, although Einstein was the intellectual architect of the experimental ideas, he did not participate in the physical execution of the experiments. Rupp did the experiments alone, meticulously following Einstein's instructions and reporting with detailed drawings and descriptions. This hands-off approach was typical of Einstein during this period, as he focused on theoretical development rather than laboratory work. However, Einstein also recognized that his collaboration with Rupp might not be well-received by Rupp's superior at Heidelberg University, the physicist Philipp Lenard, a well-known anti-relativist. Consequently, Einstein decided against conducting the experiments directly with Rupp \cite{Don}. Instead, he provided Rupp with meticulous and carefully prepared instructions, corrections, and detailed schematic drawings for each stage of the experimental process.

Moreover, it is notable that Einstein, known to often discard letters and technical correspondence from many physicists, colleagues, and friends,  conspicuously preserved Rupp's letters and diagrams. This indicates the significance Einstein attached to these experiments, which he believed had the potential to validate his commitment to wave mechanics over matrix mechanics empirically. The fact that Einstein kept these letters points to his deep interest in Rupp's results, perhaps seeing them as crucial evidence in his philosophical and scientific battle against the rising dominance of the Copenhagen interpretation.

Einstein proposed to Rupp that the results of the new experiments be published collaboratively in the \emph{Proceedings of the Prussian Academy of Sciences}. He acknowledged, however, that this joint publication might not be well-received by Rupp's superior, Lenard. To mitigate potential conflict, Einstein suggested an alternative approach: publishing their findings separately, but in consecutive order within the same issue. When Rupp provided Einstein with a final set of experimental results, Einstein facilitated their publication in the \emph{Proceedings of the Prussian Academy of Sciences}. The papers were published consecutively: Rupp's experimental results appeared alongside a theoretical paper by Einstein, in which he cited Rupp's work as empirical support for the theory.\footnote{\cite{CPAE15}, Doc. 231, 233, 240, 252, 259, 262, 270, 275, 277, 279, 285, 290, 291, 296, 299, 306, 313, 315.} 

This collaboration symbolized Einstein's enduring quest to anchor quantum mechanics in deterministic wave phenomena and his resistance to the probabilistic interpretation championed by Heisenberg and Bohr. Rupp's data, which empirically supported wave coherence and stability over long path differences, remained an essential, though separate, testament to Einstein's vision of quantum reality.

\subsection{A Sad and Cynical Twist: Rupp Pulls Einstein's Leg}

The experimental success that Einstein eagerly celebrated was, in fact, a grand deception. Rupp's canal ray experiments, which Einstein believed provided empirical support for wave mechanics, were later exposed as scientific fraud. The interference patterns, the reduction of Doppler broadening, and the supposed indirect validation of Schrödinger's wave mechanics were entirely fabricated to match Einstein's expectations. In his deep conviction of the wave nature of light, Einstein was unknowingly deceived by Rupp's skillful manipulation of data \cite{Don}.

This episode embodies a poignant irony in the historical narrative—a misjudgment that can be aptly characterized by the very adage Einstein employed to describe his ill-fated static cosmological model, the construction of a grandiose castle in the air, a vision untethered from empirical reality. 

Einstein, who fought tirelessly for realism and empirical validation, unknowingly built part of his argument against the Copenhagen interpretation on false experimental foundations. Einstein supported Rupp during the late 1920s, largely because Rupp's experiments appeared to confirm aspects of wave-particle duality that Einstein had long advocated. Rupp's experiments seemed to show interference patterns with particles (canal rays and electrons), which resonated with Einstein's ideas on the quantum nature of light and matter. Had Rupp's experiments been genuine, Einstein would have had experimental proof that wave-like interference can persist without continuous waves. This would have been a remarkable bridge between his photon theory and Schrödinger's wave mechanics, but constructed on probabilistic geometry rather than physical waves. Einstein would have argued that interference patterns can form even when light is emitted as discrete particles, not continuous waves. His so-called "ghost field" \footnote{Wave field \cite{CPAE14}, Doc. 385, section 9.}  would provide the probabilistic geometry for where those particles would interfere, even without a real physical wave present. The interference would be an effect of the geometry of the field, not the wave itself carrying energy. If Rupp's experiments had been real, they would have indirectly supported Schrödinger's view of interference as a wave phenomenon, but through a different mechanism—probabilistic geometry rather than physical waves. 
Paradoxically, Rupp's experiments would have given empirical support to the wave aspects of Schrödinger's mechanics, not by invoking continuous waves, but by demonstrating that interference is a structural property of space-time geometry. 

Buchwald and Gordin omit any hint of Rupp's fraud, framing the collaboration as a genuine scientific partnership aimed at experimental proof of Einstein's theoretical convictions \cite{BG}. 
Despite this dark twist, the Einstein-Rupp collaboration remains historically significant, not just as a tale of scientific deception, but as a window into Einstein's unwavering commitment to wave mechanics and his vulnerability to the empirical "confirmation" of his philosophical convictions.

\subsection{The Myth of Einstein's Stubbornness and Outdatedness}

According to Buchwald and Gordin, discussions around quantum mechanics during the 1920s and 1930s were not solely adversarial but often collaborative. Physicists like Paul Ehrenfest played significant roles in bridging conceptual gaps between Einstein, Bohr, and other leading thinkers of the time (they write, "Paul Ehrenfest kept trying to bring Einstein and Bohr into conversation" and "It was during this visit [at Ehrenfest's house] that Ehrenfest took the now-famous photographs of Einstein and Bohr...")\cite{BG}. The binary framing of Einstein versus Bohr conceals these cooperative intellectual exchanges that contributed to the maturation of quantum mechanics. The historical simplification into a two-person conflict neglects the broader scientific community's role in negotiating the philosophical implications of quantum theory \cite{BG}.

However, a comparative analysis of Ehrenfest's and others' influential roles in general relativity and quantum mechanics could provide a more comprehensive portrayal of Einstein's multifaceted scientific persona. 
For example, Ehrenfest played a pivotal role as an intellectual mediator in Einstein's engagement with Friedmann's expanding universe model. When Einstein visited Leiden, he often stayed with Ehrenfest, participating in lively seminars and discussions that created a fertile environment for critical reflection and exchange. During one of these visits, Yuri Krutkov, also present at Ehrenfest's residence, informed Einstein of Friedmann's letter, highlighting an error in Einstein's critique of Friedmann's work. This moment of intellectual mediation, facilitated by Ehrenfest's salon-like gatherings, proved instrumental in prompting Einstein to reevaluate his position and acknowledge Friedmann's dynamic solutions as valid extensions of his field equations \cite{Ein23}, \cite{Trop}.
Thus, Ehrenfest's role transcended that of a mere host; he served as a vital conduit for scientific dialogue, bridging gaps in communication and fostering critical exchanges that shaped the trajectory of Einstein's understanding of cosmological models.

The authors further argue that the portrayal of Einstein as "rejecting the scientific community’s consensus," "outdated," "with the relics of the past", and "reluctant to embrace modernity" is a carefully constructed myth, not a genuine reflection of his intellectual standing. This narrative was primarily engineered by the Copenhagen school and reinforced through historical accounts that positioned Bohr as the visionary of quantum mechanics while relegating Einstein to the role of a nostalgic holdover from classical physics. This is a classic example of Winner’s History—a selective historical framing that casts Bohr as the embodiment of scientific progress and Einstein as a relic of a bygone era.
In the myth of stubbornness and outdatedness construction, Einstein is portrayed as clinging to classical ideas such as determinism, realism, and separability—concepts and principles that quantum mechanics supposedly rendered obsolete. His reluctance to accept the indeterminism of the Copenhagen interpretation is framed as intellectual rigidity. At the same time, Bohr is elevated to a modern sage who transcended classical limitations and embraced the epistemological rupture that quantum mechanics demanded. This myth serves two strategic purposes: 1. It elevates Bohr's interpretation as the natural and correct progression of physics. 2. It delegitimizes Einstein's critiques, casting them as outdated, reactionary, and philosophically naïve \cite{BG}.

Buchwald and Gording argue that demythologizing this narrative reveals that it is not an organic reflection of Einstein's philosophical or scientific standing; instead, it is a construct designed to solidify Bohr’s interpretation as the definitive view of quantum mechanics. The so-called "debate" between Einstein and Bohr, immortalized in historical records, is a retrospective dramatization. It was part of a broader community dialogue that Bohr's followers historicized through their institutional influence into a mythic clash. This selective framing enhanced the narrative of Bohr's victory while simplifying Einstein's critiques into mere resistance to progress \cite{BG}.

Einstein's stubbornness is hardly a trait that bloomed late in his career. Like his well-documented debates with Bohr, Einstein engaged in intellectual confrontations with Willem de Sitter, Tullio Levi-Civita, and others, where his characteristic tenacity was prominently displayed. Throughout these debates, Einstein exhibited a steadfast commitment to the core mathematical structures underpinning his "Entwurf" theory, his solutions to gravitational waves, and his cosmological model. These exchanges were not merely technical discussions; they were interwoven with Einstein's hallmark wit, thought experiments, and innovative adjustments to his theoretical formulations, aimed at convincing his interlocutors of the soundness of his ideas.
Einstein's resistance to abandoning the core tenets of his gravitational and cosmological models mirrors his later engagements with quantum mechanics.

The reader cannot be persuaded that it was solely the Bohr school that engaged in what could be termed "historical outcome bias," notably when Einstein's debates with colleagues during the development of general relativity are omitted. 
Historiographical focus on Einstein's quantum debates, without parallel attention to his profound disagreements with colleagues during the formulation of general relativity, the prediction of gravitational waves, and the construction of his cosmological models, perpetuates this asymmetry. 
Suppose Einstein had succeeded in replacing quantum mechanics with a deterministic unified field theory. In that case, he might have been remembered as a visionary who identified deep flaws in quantum theory that others overlooked, rather than as a stubborn relic of classical physics.

An illustrative case of Einstein's dogged persistence emerges in his exchanges with the eminent mathematician Tullio Levi-Civita. Their correspondence revolved around Einstein's proof that the "Entwurf" gravitational tensor was covariant for adapted coordinate systems. Levi-Civita, armed with sharp mathematical acumen, challenged Einstein's proof repeatedly, providing detailed counterexamples that pointed to its limitations. Einstein, characteristically resolute, responded by attempting to salvage his argument, replying to Levi-Civita's critiques with a cascade of letters that circled back to the same core justifications. Einstein's unwillingness to concede was palpable; in most letters, he returned to the same reasoning, polished with minor adjustments but fundamentally unaltered.

When Levi-Civita escalated his objections, Einstein's replies grew more impassioned, expressing a blend of concern and unyielding confidence. "When I saw that you are directing your attack against the theory's most important proof, which I had won by sweat, I was a little alarmed, especially since I know that you have a much better command of these mathematical matters than I," Einstein confessed. "Yet after thorough considerations, I believe to uphold my proof right" \cite{CPAE8}, Doc. 60. At this juncture, Einstein remained committed to his adapted coordinate systems, perhaps partly out of stubbornness and genuine conviction.

Levi-Civita, however, was not so easily deterred. He continued to press Einstein on the fragility of his proof, prompting Einstein to respond with a somewhat exasperated tone, "Once again to the objection," as if Levi-Civita's insistence was merely another hurdle to clear \cite{CPAE8}, Doc. 66. Switching strategies, Einstein attempted a fresh line of reasoning to counter Levi-Civita's critique. But Levi-Civita persisted, demonstrating yet another specific case where the proof crumbled. Einstein, resolute as ever, clung to what remnants of his proof he could salvage, maintaining that Levi-Civita had merely identified a "special case" and that his critique did not undermine the theorem's general validity. "It proves nothing about the validity of the theorem in general," he wrote confidently \cite{CPAE8}, Doc. 71, undeterred by Levi-Civita's persistent rebuttals. Einstein even cheerfully informed Levi-Civita: "You have received my letter refuting your example. I shall repeat myself" \cite{CPAE8}, Doc. 69, a declaration that would make even Sisyphus pause.

Despite the mounting evidence against his position, Einstein's conviction only seemed to deepen. In one letter, he wrote with characteristic defiance, "I must even admit that, through the deeper considerations to which your interesting letters have led me, I have become even more firmly convinced that the proof of the tensor character of the 'Entwurf' gravitational tensor is correct in principle" \cite{CPAE8}, Doc. 71. One imagines Levi-Civita shaking his head, perhaps with a smile, at Einstein's unyielding resolve.

Curiously, all but one of Levi-Civita's letters were lost, while Einstein's were preserved—a typical phenomenon given Einstein's habit of discarding letters after reading them. The sole survivor of Levi-Civita's contributions was saved only because Einstein himself enclosed it back in his response, writing to Levi-Civita, "so that I can refer to it without any inconvenience to you" \cite{CPAE8}, Doc. 69. One can almost picture Einstein grinning as he penned those words, leaving a trace of his playful stubbornness for posterity.

\subsection{Rewriting the Legacy of the Einstein–Bohr Debate}

The legacy of the Einstein–Bohr debate, as it is popularly understood, is deeply entwined with what has been described by Buchwald and Gordin as \emph{Winner’s History}—the notion that the victors in scientific debates also write its history. Bohr's interpretation of quantum mechanics won in the historical narrative, with the Copenhagen school solidifying its dominance by framing Einstein's critiques as outdated and philosophically naive. "Einstein himself did not have an equivalent 'school' or group of acolytes" \cite{BG}. The shift of scientific power to the United States after World War II and the growing emphasis on practical physics over philosophical inquiry further marginalized Einstein's critiques, aligning with the notion that Bohr had effectively settled the debate.

This dominant narrative was challenged in the 1960s with John Bell's theorem, which provided a framework to test the separability principle that Einstein had championed experimentally. Despite the initial lack of attention, experiments led by John Clauser, Alain Aspect, and Anton Zeilinger from the 1970s to the 1990s eventually demonstrated quantum entanglement, confirming quantum mechanics' predictions and casting doubt on Einstein's views of locality and determinism. Ironically, it was only after these empirical confirmations that Einstein's critique was tested and refuted long after Bohr declared “victory.”
The authors argue that the experimental breakthroughs of Clauser, Aspect, and Zeilinger, culminating in the 2022 Nobel Prize in Physics, are often cited as the final proof of Einstein's “error.” Yet, this conclusion oversimplifies Einstein's legacy. Though empirically refuted in certain respects, his challenges to quantum mechanics pushed the boundaries of philosophical inquiry. They inspired new generations of physicists to probe deeper into the nature of reality. The narrative that Einstein was "left behind" is, in fact, a historical artifact that conceals his critical contributions to the foundational debates of quantum theory \cite{BG}. 

It is crucial to emphasize that although Bohr's Copenhagen interpretation, along with other interpretations of quantum mechanics that accept entanglement and its inherently probabilistic nature—such as the Many-Worlds interpretation, Quantum Bayesianism, and Relational Quantum Mechanics—has largely dominated the philosophical discourse on quantum theory, Einstein's special and general relativity profoundly transformed—and continue to transform—our understanding of space, time, and the fundamental structure of the universe.

Far from being a relic of classical physics, Einstein stands as a foundational architect of modern theoretical physics, his insights permeating even the most advanced frontiers of quantum theory and field dynamics. The birth of Quantum Field Theory (QFT) is a testament to this legacy. Emerging from the unification of quantum mechanics with Einstein's special relativity, QFT represents the synthesis of quantum principles with the relativistic requirement that the speed of light remains constant across all inertial frames. This framework would not have been possible without the conceptual scaffolding provided by Einstein's 1905 theory.

Further exemplifying Einstein's enduring influence is Feynman's path integral formulation of quantum mechanics, which relies fundamentally on relativistic invariance. Path integrals extend quantum mechanics to account for every possible trajectory a particle could take, all of which respect the constraints of special relativity. This method not only reshaped quantum theory but also became the backbone of quantum electrodynamics (QED) and modern particle physics.

Einstein's legacy is equally evident in quantum entanglement and the EPR paradox, which he helped to pioneer. Even though his original intent was to challenge the validity of quantum non-locality, his formulation of the protocol remains foundational. Moreover, despite the instantaneous collapse of entangled states, the inability to transmit usable information faster than light preserves the relativistic structure Einstein established. 

Even quantum teleportation, which leverages entanglement to transfer quantum states, adheres to this relativistic constraint, demonstrating that Einstein's influence extends far beyond classical quantum mechanics. This is poignantly reflected in Einstein's own words from his \emph{Autobiographical Notes}, where he muses on the implications of entanglement: "One can escape from this conclusion... by... assuming that the measurement of $S_1$ (telepathically) changes the real situation of $S_2$" \cite{Ein49}. 
Charles H. Bennett and his IBM Thomas J. Watson Research Center colleagues introduced quantum teleportation in their seminal 1993 paper, "Teleporting an Unknown Quantum State via Dual Classical and Einstein-Podolsky-Rosen Channels." In their introduction, the authors acknowledged the foundational role of EPR correlations in enabling long-range quantum entanglement. Bennett and his co-authors referenced Einstein's terminology, invoking his use of the word telepathically in his \emph{Autobiographical Notes}.
In their formulation, teleportation of an unknown quantum state is achieved by combining entangled EPR pairs and classical communication channels. But the protocol also strictly adheres to the no-faster-than-light constraint of special relativity \cite{Ben}.\footnote{Quantum teleportation does not allow for faster-than-light information transfer. It only transfers the state of a particle, not the particle itself, and cannot be used to send information faster than light. This maintains consistency with special relativity.}

Einstein's contributions to physics did not halt with quantum debates. Instead, they catalyzed the most profound advances in our understanding of the universe. General relativity, Einstein's magnum opus, has not only stood the test of time but has expanded in scope and sophistication far beyond what was imaginable during his lifetime. The post-1955 era saw general relativity evolve into a cornerstone of modern physics. The detection of gravitational waves by LIGO in 2015, the stunning Event Horizon Telescope (EHT) image of a black hole in 2019, and the verification of the frame-dragging effect through the satellite mission Gravity Probe B are all testaments to the enduring and expanding relevance of Einstein's theory.

Moreover, general relativity is indispensable to technologies that shape our daily lives. The GPS, for instance, would be hopelessly inaccurate if not for Einstein's relativistic corrections. Einstein's equations are crucial for satellite navigation, deep space exploration, and even simulations of the large-scale structure of the cosmos. Numerical methods and computer models of black hole collisions and neutron star mergers—made possible only through the framework of general relativity—are at the frontier of modern astrophysics.

\section{Einstein’s Political and Cultural Identity}

Exploring Einstein's political and cultural identity extends the historiographical approach established in the first chapter: demythologizing certain aspects of his legacy while recognizing how his political and cultural narrative has been mythologized. While the scientific myth casts Einstein as an isolated revolutionary in Bern and a detached philosopher in Princeton, his political and cultural life has similarly been subjected to layers of symbolic interpretation. Einstein is not only portrayed as a scientific icon but also as a global political and moral figure, symbolically claimed by various nations and movements to embody ideals of pacifism, Zionism, socialism, and intellectual freedom. This mythic framing positions Einstein as a citizen of the world, transcending borders and ideologies with seemingly universal moral clarity. However, Buchwald and Gordin demythologize this perception by contextualizing his political stances within the concrete historical realities he faced \cite{BG}.

They portray Einstein's twice-abandoned German citizenship, his affiliations with Switzerland, Italy, and the United States, and his ideological commitments—Zionism, pacifism, and civil rights advocacy—not as abstract idealisms but as practical responses to geopolitical upheavals and rising antisemitism. His political engagements are depicted as shaped by lived experiences rather than purely symbolic or ideological gestures. This perspective situates Einstein within the political realities that influenced his choices, while still acknowledging his deep personal convictions and humanitarian concerns \cite{BG}.

\subsection{Global Icon and Cultural Myth: Constructed Symbolism and Realpolitik}

Buchwald and Gordin depict Einstein as a global icon embraced by Germany, Switzerland, the United States, and Israel, alongside cities such as Ulm, Bern, Zurich, Berlin, Leiden, Prague, New York, Princeton, Pasadena, and Washington, D.C. His image was symbolically co-opted to fit various political narratives, transforming him into a symbol of pacifism, Zionism, intellectual freedom, and anti-fascism. His cosmopolitanism is reflected in his shifting national affiliations and his reluctance to be claimed exclusively by any state. His rejection of German citizenship twice and his eventual resettlement in Princeton demonstrate his disconnection from nationalist sentiments and his commitment to broader humanistic ideals.

Historiographically, the narrative dismantles the myth of Einstein as merely a world citizen by highlighting how his political stances were responses to concrete events—World War I, Nazi militarization, and the rise of McCarthyism. His support for the Hebrew University in Jerusalem, advocacy for Brit Shalom, and correspondence with Arab intellectuals underscore a pragmatic approach to Zionism that went beyond symbolic gestures to real political commitments \cite{BG}.

\subsection{Political Engagement and Moral Philosophy: From Pacifist to Realist}

The authors trace Einstein's political evolution from staunch pacifism during World War I to reluctant realism during World War II. His initial opposition to militarism, symbolized by his reaction to the Manifesto of the Ninety-Three and his support for pacifist organizations, contrasts with his decision to sign the letter to President Roosevelt urging the development of nuclear weapons—a decision driven by fear of Nazi militarization. This shift is portrayed not as a contradiction but as an adaptation to geopolitical threats that forced Einstein to reassess his stance on violence and state power.

Buchwald and Gordin position Einstein’s support for Zionism as rooted in cultural and intellectual renewal rather than blind nationalism. His involvement with the Hebrew University and his advocacy for peaceful coexistence in Palestine illustrate a vision for bi-national cooperation, diverging from mainstream Zionist politics. His rejection of the presidency of Israel in 1952 is reframed as a reflection of his ideological divergence from Israeli leadership, not merely symbolic humility \cite{BG}.

\subsection{The Sage and the Politician: From Myth to Historical Reality}

The political narrative challenges the myth of Einstein as a detached moral philosopher by illustrating his active engagement with political causes—Zionism, anti-fascism, civil rights, and global disarmament. His involvement in the atomic bomb project, his postwar activism for world government, and his critique of American militarism are recontextualized as grounded in historical reality, not abstract idealism. FBI surveillance during the Red Scare exemplifies how American liberalism selectively framed Einstein as a benign philosopher, downplaying his critiques of capitalism and militarism \cite{BG}.

This reframing of Einstein's political journey aligns with Buchwald and Gordin's broader critique of historiographical myths. Just as the binaries of Young Genius and Old Sage were dismantled, Einstein's political and cultural legacy is presented as a product of historical realities, shaped by displacement, persecution, and ideological struggle rather than mythic abstraction. His activism is depicted as historically contingent, deeply connected to real-world events, and reflective of his moral and political philosophy.

This discussion belongs under The Winners' Historiography and Political Symbolism. It reflects how Einstein's political and cultural identities were shaped and mythologized to fit dominant political narratives, mirroring the same historiographical strategies seen in the Einstein vs. Bohr debates.

\section{Free Creations of The Human Mind}

Einstein's notion of "free creations of the human mind" represents a foundational aspect of his epistemological stance. According to Einstein, the fundamental concepts and structures that underlie scientific theories are not purely derived from empirical observations or direct experiences. Instead, they are conceptual inventions, constructed by human intellect to make sense of the physical world.

Buchwald and Gordin explain that Einstein first publicly articulated this idea during his 1921 lecture, "Geometry and Experience," where he used it to describe the axioms of geometry \cite{Ein21}. He argued that while empirical data serve to test and verify these theoretical constructs, the core principles themselves are not necessarily drawn directly from experience. Instead, they are conceptual frameworks that scientists create to interpret and organize empirical data meaningfully \cite{BG}.
The authors engage with the philosophical implications of Einstein's Herbert Spencer Lecture at Oxford, \emph{On The Method of Theoretical Physics} (1933) \cite{Ein33}, situating it within its broader historical context. They observe that Einstein's lecture aligns with long-standing philosophical debates concerning the relationship between empirical evidence and reasoned reflection. By tracing these questions back to the philosophical traditions of Plato and Aristotle, the authors provide a philosophical contextualization of Einstein’s views, highlighting his contribution to the discourse on the nature of scientific knowledge. Furthermore, they emphasize Einstein's personal and political circumstances in 1933, including his forced departure from Germany and the imminent dangers he faced. In this context, the lecture is framed as a symbolic "valediction" to Europe, marking a geographical transition and an intellectual shift as Einstein moved further into his pursuit of unified field theories \cite{BG}.

In his 1933 Herbert Spencer Lecture at Oxford, Einstein distinguishes between concepts that are "free inventions of the human mind" and those that are directly derived from empirical experience. This philosophical position challenges the traditional Newtonian view that the basic principles of physics emerge solely from empirical observation. Einstein begins by tracing the development of theoretical methods back to ancient Greece, celebrating the logical rigor of Euclidean geometry as a triumph of pure reason. He describes it as an "intellectual miracle"—a logical system where propositions follow with such strict necessity that doubt seems impossible. This achievement, according to Einstein, laid the groundwork for scientific confidence and theoretical exploration. Yet, he is quick to assert that while logic provides structure, it is insufficient for comprehending the physical world. Einstein credits Galileo for recognizing that "pure logical thinking can give us no knowledge whatsoever of the world of experience; all knowledge about reality begins with experience and terminates in it." This insight, Einstein argues, marks the true birth of modern physics \cite{Ein33}.

But Einstein’s primary philosophical contribution in this lecture emerges when he examines the nature of theoretical concepts in physics. He outlines the composition of theoretical physics as consisting of "concepts and basic laws to interrelate those concepts and of consequences to be derived by logical deduction." These elements, though logically organized, are not simply abstractions from empirical data. Rather, they are, as he describes them, "free inventions of the human mind." By this, Einstein means that the foundational principles of physics—like space, time, and force—are not strictly derived from experience. They are creative constructs formulated to structure and interpret empirical observations. In this sense, theoretical physics is analogous to Euclidean geometry: both are systems of thought that impose structure upon experience through logically coherent but creatively chosen axioms.

Einstein contrasts this understanding with Newton’s perspective. Newton, whom Einstein regards as the first to create a "comprehensive and workable system of theoretical physics," believed that the foundational principles of physics were directly derived from experience. His famous assertion, "hypotheses non fingo," implied that he framed no speculative hypotheses beyond what could be observed and measured. Einstein notes that Newton and his successors viewed concepts such as space, time, mass, acceleration, and force as directly reflective of physical reality. For Newton, these were not speculative constructs but observable aspects of nature. 
Nevertheless, Einstein identifies a discomfort in Newton's writings—an unease with the notion of "absolute space" and "action at a distance." Despite this, the empirical success of Newton's theory suppressed deeper philosophical scrutiny of its foundational assumptions \cite{Ein33}.

According to Einstein, the advent of the general theory of relativity exposed the "fictitious character" of Newtonian principles. Relativity demonstrated that entirely different foundational assumptions—rooted in the concept of spacetime rather than absolute space and time—could account for empirical phenomena with even greater accuracy and coherence. This revelation, Einstein argues, illustrates that the axioms of physics are not forced upon us by nature; they are selected for their explanatory power and internal consistency. Two entirely distinct sets of theoretical foundations—Newtonian mechanics and general relativity—could both correspond to empirical reality, underscoring the idea that these principles are, indeed, "free inventions of the human mind."

Einstein elaborates that the choice of theoretical concepts is guided not by experience alone but by "mathematical simplicity and coherence." He claims that mathematics is "the truly creative principle" in theoretical physics, capable of revealing the deep structure of nature. While empirical data test the validity of these mathematical structures, they are not the source from which theoretical concepts are derived. This, for Einstein, is a defining feature of theoretical physics: it is an intellectual endeavor that transcends mere abstraction from experience, venturing instead into creative conceptualization \cite{Ein33}.

Thus, Einstein's lecture redefines the foundations of physics as creative constructions rather than purely empirical abstractions. He challenges the Newtonian belief in the direct derivation of physical concepts from experience, proposing that theoretical frameworks are intellectual inventions shaped by mathematical reasoning and validated by empirical consistency. This philosophical shift not only repositions the role of theory in science but also underscores the power of human creativity in unveiling the structure of physical reality.

Buchwald and Gordin also present Einstein's position as a counterpoint to neo-Kantianism and logical empiricism. Neo-Kantians tried to reconcile Einstein's theories with Kant's idea of space and time as a priori (pre-existing mental categories). At the same time, logical empiricists, like those in the Vienna Circle led by Moritz Schlick, believed that scientific concepts should be directly tied to empirical data. Einstein distanced himself from both, asserting that while experience tests the serviceability of scientific concepts, it is not the source from which they are derived. Instead, the truly creative principle in scientific theory lies in mathematics and the conceptual construction of models that attempt to represent reality \cite{BG}.

Einstein's notion of "free creations of the human mind" challenges the positivist dictum that theoretical terms must be reducible to direct observations. Yet, paradoxically, his insistence on empirical confirmation and his methodological commitment to experiment-driven modifications of theory invite comparisons to the logical positivist insistence on observational verification. 
This conceptual tension underscores the necessity for a deeper examination of the epistemological bridges and fault lines between Einstein's scientific realism and the logical positivist framework. 

Buchwald and Gordin's final chapter is structured more as a historical account than a rigorous philosophical analysis. They reflect on Einstein's philosophy but do not critically analyze it. They situate Einstein's epistemological reflections within significant moments of his life—his Spencer Lecture in Oxford in 1933, his interactions with the Vienna Circle, and his debates with figures like Moritz Schlick and Ernst Mach. The narrative emphasizes how Einstein's ideas evolved alongside his personal and political circumstances, notably his departure from Nazi Germany and his eventual settlement in Princeton. While not deeply analytical, Buchwald and Gordin position Einstein's epistemological stance within the two major philosophical traditions (1. neo-Kantian interpretations; 2. Mach's radical empiricism and Schlick's verificationism). The authors align Einstein’s philosophical reflections with his broader cultural and political identity. They subtly suggest that Einstein's intellectual independence—his capacity for "free creation"—mirrored his political stance as a global citizen unbound by national borders.

In the book's structure, Buchwald and Gordin strategically conclude with Einstein's notion of "free creations of the human mind" as a philosophical bridge that reframes the demythologizing narrative established in the earlier chapters. 
Throughout the book, they dismantle the myths of Einstein as the lone genius of Bern and the old sage of Princeton, emphasizing instead his deep integration within intellectual networks, collaborative dialogues, and social contexts that shaped his scientific innovations. This demythologizing project reveals Einstein not as a solitary revolutionary who invented theories in isolation, a stubborn relic clinging to outdated ideas, but as a thinker whose creativity thrived in dialogue with others. 
However, the closing chapter transcends this historical contextualization by shifting the focus to Einstein's epistemological philosophy, where scientific theories are conceptualized as "free creations of the human mind." This transition is a philosophical bridge that ties Einstein’s collaborative intellectual life to his unique creative process. By emphasizing that scientific theories are imaginative constructions rather than mere abstractions from empirical data, the authors argue that Einstein's true genius lay in his ability to invent conceptual frameworks that redefined reality, which were tested against experience. 
This philosophical lens elevates Einstein's work beyond its historical scaffolding, asserting that his theoretical breakthroughs were not just products of social networks but were acts of intellectual creativity that transcended mere empiricism. 

Thus, the closing chapter reclaims Einstein's intellectual agency, portraying him not as a passive product of his environment but as a thinker whose creative autonomy defined his scientific legacy. In this way, "free creations of the human mind" is a scaffolding concept that supports the understanding of Einstein’s process and a historiographical assertion that his imaginative constructs were essential to his revolutionary impact.


\begin{thebibliography}{17}

\bibitem [1] {Bel} Beller, M. (1999). \emph{Quantum Dialogues, the Making of a Revolution}. Chicago: University of Chicago Press. 

\bibitem[2]{Ben}  Bennet, C. H.,  Brassard, G.,  Crépeau C., Jozsa, R., Peres A., and  Wootters, W. K. (1993). “Teleporting an unknown quantum state via dual classical and Einstein-Podolsky-Rosen Channels.” \emph{Physical Review Letters} 70, pp. 1895-1899.

\bibitem[3]{Bor} Bohr, N. (1949). "Discussion with Einstein on Epistemological Problems in Atomic Physics." In Schilpp, P. A. (ed.) (1949). \emph{Albert Einstein: Philosopher-Scientist.} La Salle, IL: Open Court.

\bibitem[4]{BG} Buchwald Kormos, D. and Gordin M. D. (2025). \emph{Free Creations of the Human Mind: The Worlds of Albert Einstein}. Offord: Oxford, 2025.

\bibitem[5]{CPAE8} \emph{Collected Papers of Albert Einstein. Vol. 8: The Berlin Years: Correspondence, 1914–1918}. Schulmann, R., Kox, A.J., Janssen, M., Illy, J. (eds.). Princeton: Princeton University Press, 2002. 

\bibitem[6]{CPAE9} \emph{Collected Papers of Albert Einstein. Vol. 9: The Berlin Years: Correspondence, January 1919–April 1920}. Buchwald Kormos, D., Schulmann, R., Illy, J., Kennefick, D., and Sauer, T. (eds.). Princeton: Princeton University Press, 2004.

\bibitem[7] {CPAE14} \emph{Collected papers of Albert Einstein. Volume 14: The Berlin Years: Writings and Correspondence, April 1923-May 1925}.  Buchwald Kormos, D., Illy, J., Kox, A. J.  Lehmkuhl, D., Rosenkranz, Z. and Nollar James, J. (eds.). Princeton: Princeton University Press, 2018.

\bibitem[8] {CPAE15} \emph{Collected papers of Albert Einstein. Volume 15: The Berlin Years: Writings and Correspondence, June 1925-May 1927}.  Buchwald Kormos, D., Illy, J., Sauer, T. and Moses, O. (eds.). Princeton: Princeton University Press, 2018.

\bibitem[9]{Don} Van Dongen, J. (2007). "The interpretation of the Einstein-Rupp experiments and their influence on the history of quantum mechanics." \emph{Historical Studies in the Physical and Biological Sciences} 37, pp. pp. 121-131.

\bibitem[10]{Ein21} Einstein, A. (1921). "Geometrie und Erfahrung." \emph{Preußische Akademie der Wissenschaften (Berlin). Sitzungsberichte} I, pp. 123-130. 

\bibitem[11]{Ein23} Einstein, A. (1923). "Notiz zu der Arbeit von A. Friedmann 'Über die Krümmung des Raumes'." \emph{Zeitschrift für Physik} 16, p. 228.

\bibitem[12]{Ein33} Einstein, A. (1933) \emph{On The Method of Theoretical Physics.} Oxford: Oxford University Press.

\bibitem[13]{Ein49} Einstein, A. (1949). “Autobiographical Notes.”  In Schilpp, P. A. (ed.) (1949). \emph{Albert Einstein: Philosopher-Scientist.} La Salle, IL: Open Court.

\bibitem[14]{Gal} Galilei, G. (1632). \emph{Dialogue Concerning the Two Chief World Systems – Ptolemaic and Copernican.} Trans. Drake, S. Berkeley: University of California, 1967.

\bibitem[15]{Hoff} Hoffmann, B. (1968). "My Friend, Albert Einstein." \emph{Reader's Digest}. January.

\bibitem[16]{Hol} Holton, G. (1969). "Einstein and the 'Crucial' Experiment." \emph{American Journal of Physics} 37, pp. 968-982.

\bibitem[17] {Jan} Janssen, M. and Renn, J. (2015). "Arch and scaffold: How Einstein found his field equations." \emph{Physics Today} 68, pp. 30-36.

\bibitem[18]{Pais} Pais, A. (1982). \emph{Subtle is the Lord. The Science and Life of Albert Einstein.} Oxford: Oxford University Press.

\bibitem[19]{Schi} Schilpp, P. A. (ed.) (1949). \emph{Albert Einstein: Philosopher-Scientist.} La Salle, IL: Open Court.

\bibitem[20]{Shan} Shankland, R. (1963). "Conversations with Albert Einstein I". \emph{American Journal of Physics} 31, pp. 47-57.

\bibitem[21]{Trop} Tropp, E. A., Frenkel, V. Y. and Chernin, A. D. (1993). \emph{Alexander A Friedmann: The Man who Made the Universe Expand.} Cambridge: Cambridge University Press.

\end{thebibliography}
\end{document}